\documentclass{PoS}

\usepackage{epsfig}
\usepackage{graphicx}

\title{Fitting and selecting scattering data}

\ShortTitle{Fitting and selecting scattering data}

\author{\speaker{Enrique Ruiz Arriola}\thanks{This work is supported
  by the Spanish Ministerio de Econom\'{\i}a y Competitividad and European
  FEDER funds under contracts FIS2014-59386-P and
  Junta de Andaluc\'{\i}a grant FQM-225}\\
Departamento de F\'{\i}sica At\'{o}mica, Molecular y 
  Nuclear and Instituto Carlos I de \\ F{\'\i}sica Te\'orica y Computacional. 
Universidad de Granada, E-18071 Granada, Spain.\\
        E-mail: \email{rnavarrop@ugr.es}}

\author{Jose Enrique Amaro\\
Departamento de F\'{\i}sica At\'{o}mica, Molecular y 
  Nuclear and Instituto Carlos I de \\ F{\'\i}sica Te\'orica y Computacional. 
Universidad de Granada, E-18071 Granada, Spain.\\
        E-mail: \email{amaro@ugr.es}}

\author{Rodrigo Navarro P\'erez\\
Institute  of  Nuclear  and  Particle  Physics,  and  Department  of
Physics  and  Astronomy, \\ Ohio  University,  Athens,  OH  45701,  USA\\
        E-mail: \email{navarrop@ohio.edu}}

\abstract{The main purpose of scattering experiments is to unveil the
  underlying structure of the colliding particles and their
  interaction. Typically one measures scattering observables (cross
  sections and polarizations) at discrete angles and energies and
  mutually consistent data may validate or falsify proposed theories
  or models. However, the accumulation of data from different
  laboratories while potentially improves the statistical significance
  it may sometimes generate mutually inconsistent data as a
  side-effect. Thus, some decision has to be made on what are the
  maximal amount of data which are mutually compatible. We show
  elastic $\pi N $ and $NN$ scattering as prominent examples where
  this selection is called for. We discuss how it can be done in a
  self-consistent manner invoking a principle of maximal consensus of
  the database and with the help of a sufficiently flexible model
  involving a minimal number of theoretical assumptions. In the NN
  case this has become possible with a combination of long distance
  field theoretical constraints at the hadronic level such as pion
  exchanges and electromagnetic effects and a coarse graining of the
  unknown interaction over the shortest de Broglie wavelength being
  probed in the scattering process.}

\FullConference{XVII International Conference on Hadron Spectroscopy and Structure - Hadron2017\\
		25-29 September, 2017\\
		University of Salamanca, Salamanca, Spain}

\begin{document}

\section{Introduction}

Fitting and selecting scattering data are intertwined activities with
a long tradition in Nuclear and Particle Physics.  In the absence of
mutually inconsistent data, theories may be validated or falsified by
data, since only in such a case a satisfactory fit becomes
possible. Thus, ensuring that this is indeed the case allows one to
answer important questions. We stress from the beginning that the
issues which we will be discussing were proposed many years
ago~\footnote{To our knowledge the first time a fit using the least
  squares method was applied to analyze $\pi N$ scattering data in
  terms of phase-shifts was in 1952 by Fermi and
  Metropolis~\cite{fermi1952alamos}. They proposed a derivative-free
  algorithm and used the MANIAC (Mathematical Analyzer, Numerical
  Integrator, and Computer), the first computer at Los Alamos designed
  by Metropolis. In the Rochester conference in 1952 (see
  \cite{schweber1955mesons} for a review) the value of $\chi^2$ was
  actually used to invalidate incompatible phase-shifts
  determinations.  Error determination of phase-shift was analyzed
  soon thereafter in 1955 by means of the well-known error
  matrix~\cite{Anderson:1955zza} which used the improved AVIDAC
  (Argonne Version of the Institute's Digital Automatic
  Computer). This example was followed to undertake a similar analysis
  in the np and pp scattering case~\cite{Stapp:1956mz}. It is
  surprising that even to this day, 70 years later, presumably
  benchmarking analyses lack this simple error estimates. The issue of
  scattering data selection was started triggered by the accumulation
  of data which at times were mutually
  inconsistent~\cite{perring1963nucleon}. This approach was followed
  by the Livermore fit (see e.g. for a
  review)~\cite{arndt1966chi}.}. In this contribution we revisit the
subject at a comprehensive level as we feel that essential issues are
too often forgotten.  More details and references within the NN
context can be traced
from~\cite{Perez:2013jpa,Perez:2014yla,Perez:2014kpa,Perez:2014bua,Perez:2014waa,RuizArriola:2016sbf,Perez:2016aol,RuizSimo:2017anp}.

\section{Scattering}

%\underline{\it Experiment}.  

Scattering experiments are designed after the original Rutherford
experiment which lead to the discovery of the atomic nucleus in 1908:
$N_{\rm in}$ particles emitted from a source of surface $S$ are
collimated forming a beam which is scattered at a given solid angle
$\Omega \equiv (\theta,\phi)$ and $N_{\rm out}$ particles are counted
on a detector at a far distance $R$. The differential cross section is
defined as the ratio
\begin{eqnarray}
\sigma_{\rm exp} (\theta,\phi) = \frac{N_{\rm out} (\theta, \phi) /\Delta \Omega}{N_{\rm in} /S} \, , 
\end{eqnarray}
over a given time interval and detector angular resolution $\Delta
\Omega =(\Delta \theta,\Delta \phi)$. In general, there exists a
normalization constant, which can be determined by comparing with a
theoretically {\it known} cross section or by checking the total cross
section $\sigma_T \equiv \int d\Omega \sigma(\Omega) $ with a forward
transmission experiment where the mean free path is determined $l = 1/
n \sigma_T $ and the density of scatterers per unit volume, $n$, is
known.  We assume for simplicity non-relativistic elastic scattering
for spinless particles interacting by a central potential $V(r)$.

%\underline{\it Classical}. 
At the classical level, one solves Newton's
equation $ \mu {\vec x}''(t) = - \vec \nabla V(\vec x)$ subjected to
the conditions $\vec x(t) \to \vec b + \vec v t $ and $\vec x(t) \to
\vec b' + \vec v' t $ for $t \to \mp \infty$ respectively, where $\vec
b \cdot \vec v= b' \cdot \vec v'= 0$, $\hat v' \cdot \hat v = \hat b'
\cdot \hat b = \cos \theta$ and $\sigma_{\rm cl} (\theta,\phi) = d^2 b
/d\Omega = (b / \sin \theta)(db/d\theta) $.  Angular momentum
conservation implies $ | \vec L (t) | = |\vec x (t) \wedge \vec p (t)
| = b p = b \mu v $ and for a finite range interaction, i.e. $V(r)=0$
for $r > a$, so that $\sigma_{\rm cl} (\theta,\phi)=0$ for $b>a$. The
meson exchange picture yields Yukawa-like forces among hadrons $ V(r)
\sim e^{-(r/a)}/r$ with the longest range corresponding to $a \sim
\hbar / m_\pi c \sim 1.4 {\rm fm}$. For these forces the total classical cross
section diverges, $\sigma_T \equiv \int d\Omega \sigma_ {\rm cl} (\Omega) = \int
d^2 b = \infty$.

%\underline{\it Quantum}. 
Quantum mechanically one has $\sigma_{\rm QM}
(\theta,\phi) = |f(\theta,\phi)|^2$ where the scattering amplitude reads 
  \begin{eqnarray}
  f(\theta,\phi) = \sum_{l=0}^\infty (2l+1) \frac{e^{2i\delta_l (p)}-1}{2ip}P_l(\cos \theta) \quad , \qquad E= \frac{p^2}{2\mu} \, . 
  \end{eqnarray}
  Here, $P_l (z)$ are Legendre polynomials and $\delta_l(p)$ are the
  phase-shifts which are computed by solving the reduced Schr\"odinger
  equation for the reduced wave function $u_l(r)$ ($\Psi(\vec x)=
  (u_l(r)/r) Y_{l,m} (\theta,\phi)$), 
  \begin{eqnarray}
  -u_l''(r) + \left[ \frac{l(l+1)}{r^2} + 2\mu V(r) \right] u_l(r) = p^2 u_l (r) 
  \end{eqnarray}
  with the asymptotic conditions (we assume  non-singular potentials $r^2  V(r) \to 0$)
  \begin{eqnarray}
u_l(r) \underbrace{\to}_{r \to 0} r^{l+1} \quad , \qquad 
u_l(r) \underbrace{\to}_{r \to \infty} \sin \left( p r - \frac{l\pi}{2}+ \delta_l \right)
  \end{eqnarray}
The no-scattering condition corresponds to take $L^2 = l(l+1)^2 \sim
(l+1/2)^2$ so that $ V(r) \sim 0 $ for $ r \gtrsim a $ corresponds to
$\delta_l (p) \sim 0 $ for $ b \gtrsim a $ or equivalently $ l_{\rm
  max} + \frac12 \sim p a \sim p/m_\pi $. In this case the total cross
section is now convergent for Yukawa forces. The (truncated) partial
wave analysis (PWA), describes scattering data in terms of
phase-shifts and {\it known} angular dependence of $P_l(\cos \theta)$.

\begin{figure}[hpt]
  \begin{center}
    \epsfig{figure=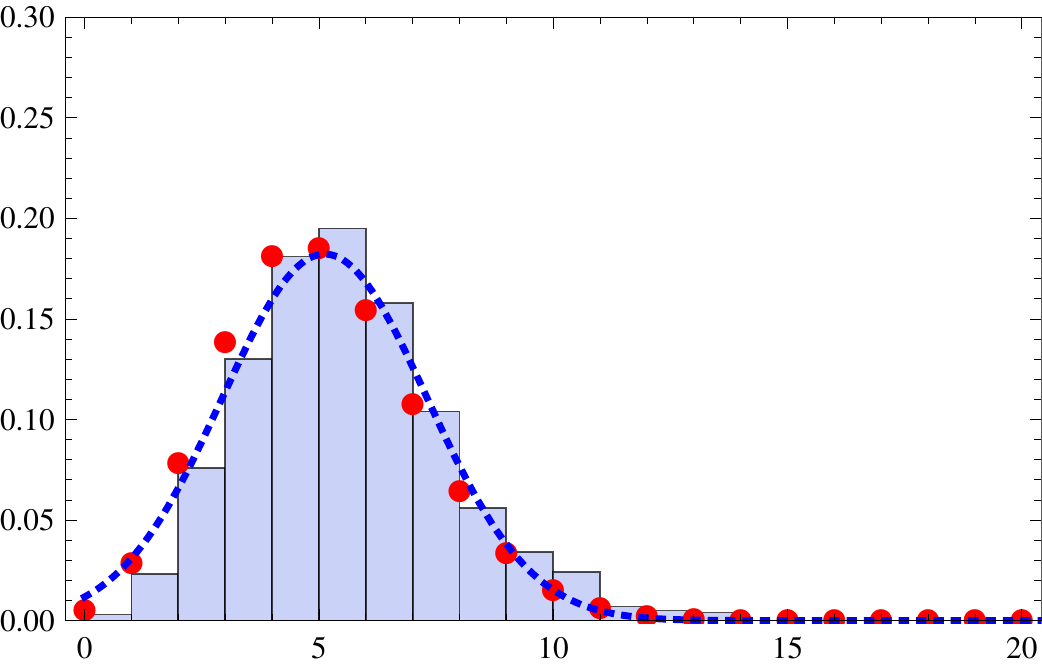,width=0.45\linewidth}
    \caption{Binomial (Histogram), Poisson (Points) and Gauss (Line)
      distributions for $p=0.1$ and $N=50$.\label{Fig:count}}
  \end{center}
\end{figure}

In general, the number of counts $N_{\rm
  out}$ is a random variable subjected to fluctuations. If for a fixed
solid angle $(\theta, \phi)$ we denote by $p$ the scattering
probability, the statistics of counting $k$ events out of $N$ is given
by the binomial distribution with mean $\bar k$ and variance $(\Delta
k)^2$, 
\begin{eqnarray}
P_{N,k} &=& 
\left(
\begin{array}{c}
 N \\
 k
\end{array}
\right) p^k (1-p)^{N-k} 
\, , \quad \bar k = N p  \, , \quad (\Delta k)^2 =
\overline{ ( k - \overline{k} )^2 } = 
\overline{k^2}  - \overline{k}^2 = N p(1-p)   \, . 
\end{eqnarray} 
In practice, $p \ll 1 \ll k \ll N$ one has the sequence of Binomial $\to$
Poisson $\to$ Gauss distributions
\begin{eqnarray}
P_{N,k} &\underbrace{\to}_{p \ll 1} 
\frac{e^{-Np} (Np)^k}{k!} \underbrace{\to}_{k \gg 1}  
\frac{e^{-(k-Np)^2/2}}{\sqrt{2\pi} \Delta k} \, , 
\end{eqnarray} 
where in Fig.~\ref{Fig:count} we illustrate the situation for the case
$p=0.1$ and $N=50$. Thus, we may consider that $N_{\rm out}$ (and
hence $\sigma (\theta,E) $) is Gauss distributed. For a $68\%$
confidence level we write as usual~\footnote{If $\xi$ is a normally 
distributed variable, i.e. $\xi \in N(0,1)$ the probability density is 
$P(\xi) = e^{-\xi^2/2} /\sqrt{2\pi}$ with 
$\bar \xi=0$ and $\Delta \xi=1$ and hence we may write 
$N_{\rm out} = \bar N_{\rm out} + \xi \Delta N_{\rm out}$.}
\begin{eqnarray}
N_{\rm out} = \bar N_{\rm out} \pm \Delta N_{\rm out} \, , \qquad \Delta N_{\rm out} = 
\sqrt{  \bar N_{\rm out}} \, , 
\end{eqnarray}

%\begin{eqnarray}
%\sigma (\theta,E) = \bar \sigma (\theta,E) \pm \Delta \sigma (\theta,E) \, . 
%\end{eqnarray}

\section{Fitting}

\begin{figure}[hpt]
\begin{center}
\includegraphics[width=0.45\linewidth]{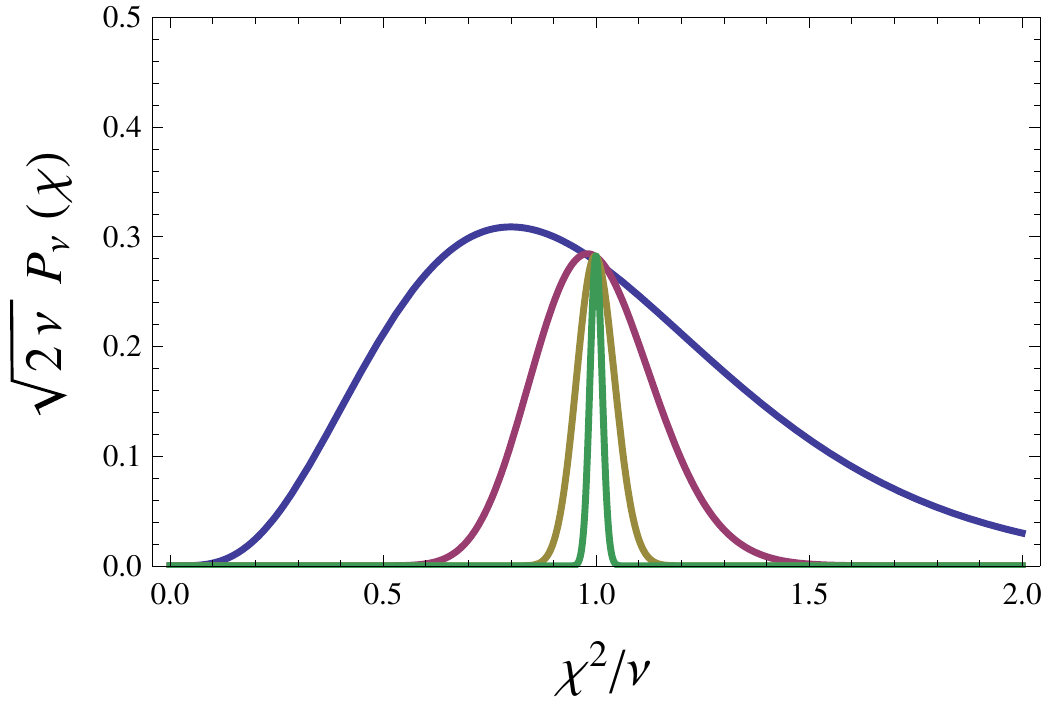}
\hskip.5cm
\includegraphics[width=0.45\linewidth]{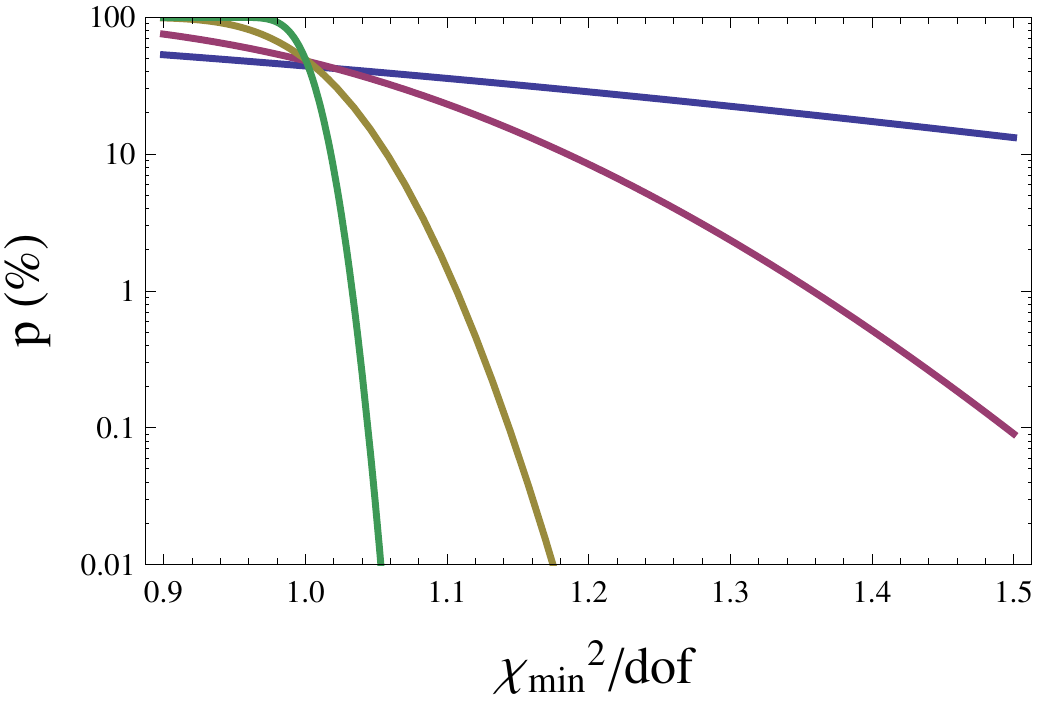}
\caption{
The $\chi^2$ distribution for $\nu=10$ (blue),100 (pink),1000 (brown),10000 (green).
Left panel: $\sqrt{2 \nu} P_\nu (\chi) $ as a function of $\chi^2 /\nu$.  Right panel: $p$-value defined as $p = 100 \times \int_{\chi_{\rm min}^2}^\infty P_\nu (\chi) d \chi$. \label{fig:p-value}}
\end{center}
\end{figure}

\underline{\it Statistics}. The general fitting problem corresponds to
a situation where we have $N$ data with uncertainties $ O_i \pm \Delta
O_i $ and a theory depending on $M$-parameters ${\bf p}= (p_1 , \dots,
p_M )$ predicting $ O_i ({\bf p}) $. The question: Does theory explain
data ? can be answered statistically as follows. If all uncertainties
follow an independent Gaussian distribution for a choice of parameters
$ {\bf p}$ we write 
\begin{eqnarray}
O_i^{\rm exp} = O_i^{\rm th} + \xi_i \Delta O_i \, ,  \qquad \xi_i \in N(0,1) \, , 
\label{eq:fit}
\end{eqnarray}
we define the minimized {\it least squares sum} 
\begin{eqnarray}
\chi_{\rm min}^2 \equiv {\rm min}_{\bf p} \chi^2 ({\bf p}) = \chi^2
({\bf p}_0) \, , \quad \chi^2 ({\bf p}) = \sum_{i=1}^N \left[
  \frac{O_i({\bf p})-O_i^{\rm exp}}{\Delta
    O_i}\right]^2 \, . 
\end{eqnarray} 
This condition effectively eliminates $M$ independent variables, so
that for the remaining degrees of freedom $\nu = N-M$ one has the
following $\chi^2$ probability density distribution
\begin{eqnarray}
P_\nu (\chi^2) = \prod_{n=1}^\nu \left( \int_{-\infty}^\infty d \xi_i \frac{e^{-\xi_i^2/2}}{\sqrt{2\pi}} \right)  
\delta ( \chi^2 - \sum_{n=1}^\nu \xi_n^2) = \frac{ e^{-\chi^2} \chi^{\nu-2}}{2^{\nu /2}\Gamma \left(\frac{\nu }{2}\right)} \, ,  
\end{eqnarray}
which is plotted in Fig.~\ref{fig:p-value} as a function of $\chi^2 /
\nu$, and shows the drastic narrowness for a large number of
data. The mean and variance are given by $\langle \chi^2 \rangle = \nu
$, $ \langle ( \chi^2 - \langle \chi^2 \rangle)^2 \rangle = 2 \nu^2
$. For $\nu \gg 1$ the $\chi^2 \in N(\nu, \sqrt{2 \nu})$ whence $
\chi^2 = \nu \pm \sqrt{2 \nu} $ at the $68\%$ confidence level.  Thus,
the assumption that data differ from theory by fluctuations,
Eq.~(\ref{eq:fit}), holds at $N_\sigma$-standard deviations level if
\begin{eqnarray}
\frac{\chi_{\rm min}^2}{\nu} =  1 \pm  N_\sigma \sqrt{ \frac{2}{\nu}} \, ,  \quad \nu = N-M \, \quad ,  {\rm d.o.f \, \, (degrees\, of \, freedom) } \, . 
\label{eq:st}
\end{eqnarray}
 Table~\ref{tab:chi2} provides the case $N_\sigma=1$. Thus, $\chi_{\rm
   min}^2/\nu $ outside the confidence interval is unlikely (for
 $N_\sigma=1,2,3$ is less than $32, 5 , 1\%$ respectively) and implies
 either a bad model or bad data or both. On the contrary, an
 acceptable $\chi^2/\nu $ suggests consistency between model and data
 and, more importantly, errors on the parameters reflect statistical
 uncertainties of the input data ${\bf p}={\bf p}_0+ \Delta {\bf p} $
 which can be propagated to functions of the parameters $F({\bf p})$
 not involved in the fitting procedure.

%A desirable property is that these parameters are
%uncorrelated~\cite{Perez:2014kpa}.

\begin{table}
  \begin{center}
  \begin{tabular}{|c|cccc|}
    \hline 
  $\nu$ &  10 &  100 &  1000 &  10000 \\  
  $   \chi^2/\nu \, $ ($ 68 \% $) &
 $1 \pm 0.447 $  &
 $1 \pm 0.141 $ &
 $1 \pm 0.044 $ &
 $1 \pm  0.014. $ \\
 \hline 
  \end{tabular}
  \end{center}
  \caption{Values of acceptable fits for some sample sizes at the $1\sigma$-level.}
\label{tab:chi2}
\end{table}

\underline{\it Single energy fits}. The simplest situation corresponds
to have complete data in a {\it given} energy $E$ (or momentum $p$),
namely $ \left(\sigma(\theta_1,E) , \dots , \sigma(\theta_N,E)
\right)$. In this case one can determine the $l_{\rm max} \sim p a$
phase-shifts {\it directly} from the data as fitting parameters $
(\delta_0 (E), \dots , \delta_{l_{\rm max}} (E)) $ by minimizing
\begin{eqnarray}
\chi^2 ( \delta_1 (E), \dots , \delta_{l_{\rm max}} (E), Z) =
\sum_{i=1}^N \left[ \frac{ \sigma^{\rm exp} (\theta_i, E) - Z
    \sigma^{\rm th} (\theta_i,\delta_1 (E), \dots , \delta_{l_{\rm
        max}} (E))}{ \Delta \sigma (\theta_i, E) } \right]^2 +
\left(\frac{1-Z}{\Delta Z} \right)^2 \end{eqnarray} Here the
normalization $Z$ with estimated uncertainty $\Delta Z$ (provided
by experimentalists) is {\it common} for {\it one} energy.
Phase-shifts become ``experimental'' and {\it model independent}
observables, $ \delta_l^{\rm exp} (E) \pm \Delta \delta_l^{\rm exp} (E) $ for 
$ l=0, \dots , l_{\rm max} $. 
%\begin{figure}
%[hpt]
%\begin{center}
%\epsfig{figure=../Calculations/Abundance.pdf,width=0.45\linewidth}
%\caption{Complete set of measurements}
%\end{center}
%\end{figure}

\underline{\it Multiple energy fits}.  If one has incomplete data in
energies and angles $ (\sigma(\theta_1,E_1) , \dots ,
\sigma(\theta_N,E_N))$ one cannot generally determine phase-shifts
$\delta_l (E_i)$ at those energies. Instead, a {\it model dependent}
interpolation with fitting parameters ${\bf p}$ in the energy is
needed. Thus, one minimizes
%Assuming smoothness
%in $(\theta,E)$ one is driven instead to fit 
\begin{eqnarray}
\chi^2 ( {\bf p} , Z) = \sum_{i=1}^N \left[ \frac{ \sigma (\theta_i,
    E_i)^{\rm exp} - Z \sigma^{\rm th} (\theta_i, E_i, {\bf p})}{
    \Delta \sigma (\theta_i, E_i) } \right]^2 +
\left(\frac{1-Z}{\Delta Z} \right)^2 
\end{eqnarray}
Different experiments have
different normalizations so that generally
\begin{eqnarray}
\chi^2 ( {\bf p} , Z_1, \dots Z_E ) = \sum_{i=1}^E \chi_i^2 ( {\bf p} , Z_i ) \label{eq:chinu}
\end{eqnarray}
%The statement 
%\begin{eqnarray}
%\sigma (\theta_i, E_i)^{\rm exp} =  \nu_0 \sigma^{\rm th}
%(\theta_i, E_i, {\bf p}_0) \pm \Delta \sigma (\theta_i, E_i) 
%\end{eqnarray}

\underline{\it Coarse graining}. In the present situation a
multienergy fit becomes mandatory and the question is how to choose
the fitting parameters ${\bf p}$. Our approach is based in separating the
potential into a field theoretical piece and a unknown coarse grained
piece at points $r_n$ suitably located
\begin{eqnarray}
V(r)  = \left[ \sum_n \Delta r V(r_i) \delta (r-r_n) \right] \theta ( r_c-r ) +
V_{\rm QFT} (r) \theta ( r-r_c ) \quad , \qquad  r_n = n \Delta r 
\end{eqnarray}
with $\Delta r \sim 1/p_{\rm max}$ the shortest de Broglie wavelength
and $r_c$ provides a boundary which should be larger than the
elementarity size of the hadrons $r_e$ and turns out to be $r_c=1.8-3
{\rm fm}$. This allows to determine {\it a priori} the number of
fitting parameters, $V(r_i)$, to be $N \sim (p r_c)^2 $, which are
determined with errors $\pm \Delta V(r_i)$ and turn out to be mostly
uncorrelated for NN in different partial waves~\cite{Perez:2014yla}.

\section{Selecting data}

\begin{figure}[hpt]
\begin{center}
\epsfig{figure=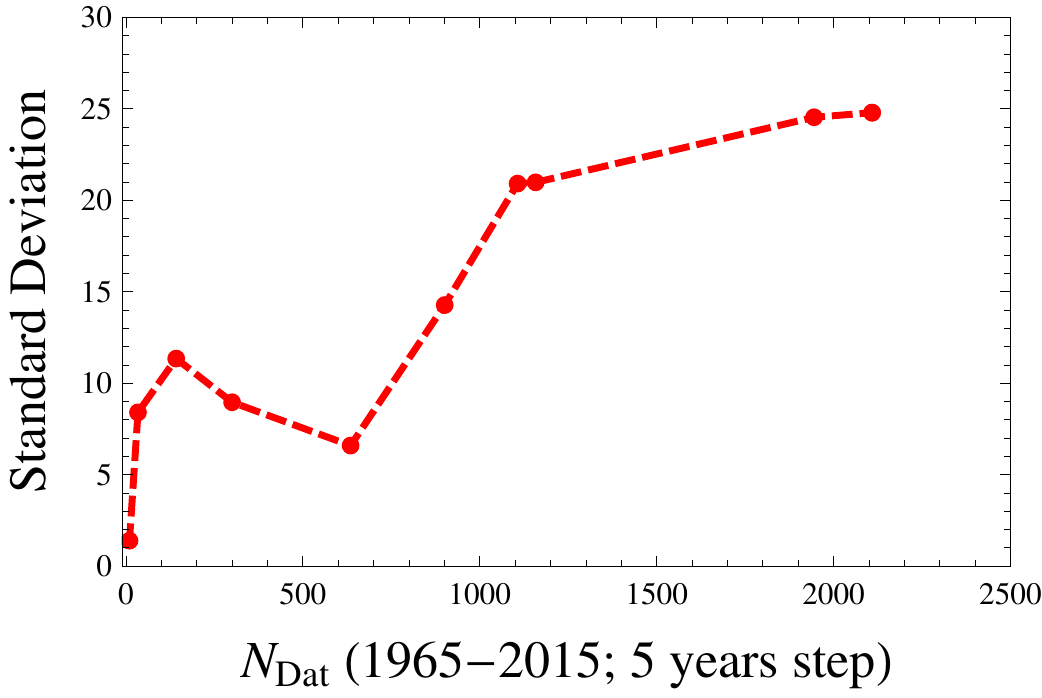,width=0.45\linewidth}
\epsfig{figure=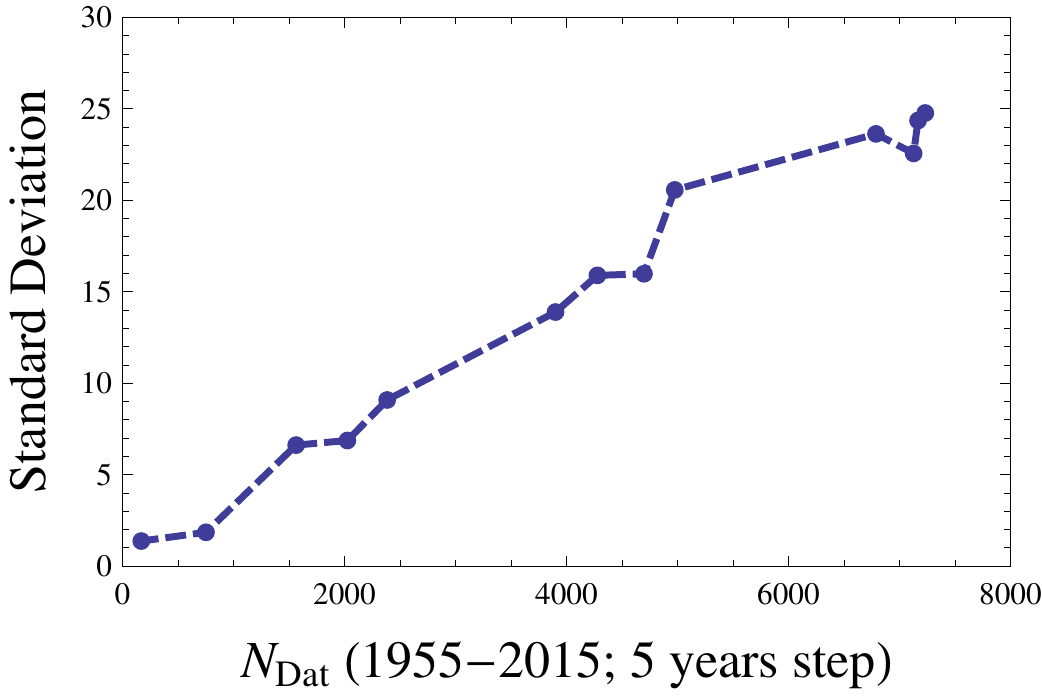,width=0.45\linewidth}
\caption{Evolution of the standard deviation $N_\sigma$ defined by
  Eq.~(\ref{eq:st}) as a function of the number of data in 5 years
  steps for $\pi N$ at $T_{\rm LAB}^\pi \le 200 {\rm MeV}$ in the
  period 1965-2015 (left panel) and $ N N = np +pp $ at $T_{\rm LAB}
  \le 350 {\rm MeV}$ in the period 1955-2015 (right panel). We use the
  SAID database~http://gwdac.phys.gwu.edu/.\label{fig:SD}}
\end{center}
\end{figure}

The accumulation of data improves the statistics and the precision but
often generates mutually incompatible data. The need for selection
becomes evident from Fig.~\ref{fig:SD} for the cases of elastic $\pi
N$ and $NN$ scattering where the considered SAID fits display unlikely
large $N_\sigma$ from Eq.~(\ref{eq:st}). Taking data is painstaking
and costs money. As theoreticians we do not feel qualified enough to
discuss what our colleagues experimentalists do in their labs,
specially since questioning one experiment means in fairness
questioning all experiments. This may not necessarily mean genuinely
wrong experiments, but rather {\it unrealistic} error estimates. Note
that the goal of a fit is to determine the true values of certain
parameters with a given and admissible confidence level. Fortunately,
for a large number of experiments we can use statistics and select
data according to a principle of consensus among experiments
arbitrated by a fitting model.  Therefore, we will assume that all
published experiments are correct unless proven otherwise and look for
the flexible enough theory which congregates the largest number of
data with an acceptable $\chi_{\rm min}^2/\nu \sim 1 \pm \sqrt{2/\nu}$. 

For NN scattering and $T_{\rm LAB} \le 350 {\rm MeV}$ we have extended
the standard $3\sigma$ criterion into a self-consistency by the
following selection process (see~\cite{Perez:2013jpa} for details): 1)
Fit to all data.  If $\chi^2/\nu \lesssim 1$ you can stop. If not, proceed
further. 2) Remove data sets with improbably high or low contributions
to the total $\chi^2$ (3 $\sigma$ criterion). 3) Refit parameters for
the remaining data.  4) Re-apply $3\sigma$ criterion to all data. 5)
Repeat until no more data are excluded or recovered.

The effect of this selection procedure is to go from $\chi^2/\nu|_{\rm
  all}=1.41 $ to $ \chi^2/\nu|_{\rm selected}=1.04$ with a reduction
in the number of data from $N_{\rm Data}= 8173 $ to $ N_{\rm Data}=
6713$.  While this seems a drastic rejection, the NN Granada-2013
database is the largest one to date providing a self-consistent fit
below 350 MeV. For this number of data this is {\it not} a minor
improvement as can be seen from Fig.~\ref{fig:p-value}. When the
pion-nucleon coupling constants become fitting parameters $
\chi^2/\nu|_{\rm selected}=1.025 $~\cite{Perez:2016aol}.  Moreover,
normalization factors in Eq.~(\ref{eq:chinu}) turn out to be very
close to unity~\cite{Perez:2013jpa} reducing the impact of an interlab
bias~\cite{d2003bayesian}~\footnote{We thank Jacobo Ruiz de Elvira for
  drawing our attention to this issue.} an issue which will be
reported elsewhere. Finally, one should keep in mind that the needed
data need not be $(E,\theta)$ distributed in a way that information is
most useful, precise and non-redundant, a relevant aspect for large
databases.  If we have many data and/or accurate data in a given
region of energies and angles they will tend to dominate the fit over
some other regions for which there exist less data. Therefore, there
is an abundance bias.

\section{Discussion and outlook}

There is a natural reluctance to discard data without going into the
intricacies of the experiment, particularly if they are produced by
influential collaborations and possibly one should simply ask the
experts. When the number of experiments is large and data have been
collected over the years there is no sound reason to discard old data
as if wrong, because any data will eventually become old at some
point. While errors are expected to be realistic, one should keep in
mind that any experiment may appear to be consistent simply because
both statistical and systematic errors are taken to be
``conservatively'' large (but also irrelevant in the $\chi^2$) and
inconsistent if errors are ``boldly'' small (and extremely
influential). A decision needs some underlying model and the best one
should provide an acceptable fit, congregating as many data as
possible.  In the present contribution we have shown that we can
invoke statistics to discard scattering data according to the
principle of maximal consensus. The coarse graining approach has
proven to be a successful framework to select NN scattering data and
hence to answer important
questions~\cite{RuizArriola:2016sbf,Perez:2016aol,RuizSimo:2017anp}. It remains to be
seen if such an approach can be extended to other hadronic systems of
interest.

%\bibliographystyle{JHEP}
%\bibliography{refs}

\providecommand{\href}[2]{#2}\begingroup\raggedright\endgroup

\end{document}